\documentclass[journal]{IEEEtran}
\usepackage{xspace,amsmath,amssymb,amsfonts,epsfig,syntonly}
\usepackage{diagbox}
\usepackage{cite,bm,color,url,textcomp}
\usepackage{epstopdf}
\usepackage{empheq}
\usepackage{graphicx}
\usepackage{subfigure}
\usepackage{ragged2e}
\usepackage{array}
\usepackage[utf8]{inputenc}

\usepackage{bbm}
\usepackage{booktabs,framed}

\usepackage{xspace,amsmath,amssymb,amsfonts,epsfig,syntonly}
\usepackage{bm,color,url,textcomp}
\usepackage{epstopdf}
\usepackage{empheq}
\usepackage{graphicx}
\usepackage{algorithm} 
\usepackage{algorithmic} 
\usepackage{multirow} 
\usepackage{xcolor}
\usepackage{makecell}

\newcommand{\cmmnt}[1]{\ignorespaces}

\long\def\symbolfootnote[#1]#2{\begingroup
\def\thefootnote{\fnsymbol{footnote}}
\footnote[#1]{#2}\endgroup}

\psfull

\allowdisplaybreaks[4]

\begin{document}
\title{
Throughput Maximization for MapReduce-Based Collaborative Computing over Energy-Harvesting Wireless Devices
}

\author{Yuhang Li, Siqi Sun, Hongen Zheng, Xiaojing Chen, Shunqing Zhang,~\IEEEmembership{Senior Member, IEEE}, and Yanzan Sun
\thanks{This work was supported by Shanghai Municipal Science and Technology Commission Foundation grant 25DP1500300. \textit{(Corresponding author: Xiaojing Chen.)}

The authors are with the Key Laboratory of Specialty Fiber Optics and Optical Access Networks, Shanghai University, Shanghai 200444, China. 
Emails:~\{liyuhang, sunsiqi1303, 13482746459, jodiechen, shunqing, yanzansun\}@shu.edu.cn.
}
}

\maketitle

\setcounter{page}{1}
\begin{abstract}
This paper studies resource allocation for MapReduce-based collaborative
computing over heterogeneous wireless devices powered by renewable energy
harvesting. We formulate a long-run average throughput maximization problem
that jointly optimizes computing load, phase time allocations, transmit
power, and per-device energy consumption, subject to battery evolution,
CPU frequency, and latency constraints. To solve this problem online without
prior knowledge of channel states or energy arrivals, we propose a
DDPG-CVX algorithm that couples Deep Deterministic Policy Gradient (DDPG)
with convex programming. DDPG determines the per-slot energy budget for each
device from observed battery and channel states; the remaining resource
allocation variables are then resolved to global optimality by an embedded
convex solver. This two-phase decomposition reduces the action-space
dimensionality of DDPG while preserving per-slot solution quality.
{\color{black}Simulations show that DDPG-CVX achieves 1.25$\times$$\sim$32.36$\times$ the throughput of representative benchmarks.}
\end{abstract}


\section{Introduction} 
Intelligent applications, such as real-time deep learning (DL) inference, extended reality, and autonomous navigation, impose heavy computational demands on resource-constrained end devices~\cite{khoshsirat2024decentralized,yang2020energy}. Collaborative computing alleviates this burden by distributing inference tasks among heterogeneous wireless devices~\cite{11240206}. The MapReduce framework is a natural fit: Each device computes partial results in the Map phase, exchanges intermediate outputs through an access point (AP) in the Shuffle phase, and aggregates them in the Reduce phase~\cite{8744394}. For renewable-energy-powered devices, energy-aware resource management is essential to sustain this collaborative pipeline.

Resource allocation in this framework faces three challenges. First, existing schemes often assume homogeneous devices with similar computing capabilities and energy budgets~\cite{8051074,paris2021leveraging}, whereas practical heterogeneity in hardware, channel conditions, and harvested energy can reduce throughput and energy efficiency. Second, prior studies mainly focus on computation--communication tradeoffs from a network-coding perspective~\cite{11005962,9795129,cheng2024asymptotically}, leaving the joint optimization of computing load, time allocation, transmit power, and energy largely unexplored. Third, finite batteries and uncertain renewable energy arrivals couple current decisions with future feasibility, making methods based on perfect channel and energy knowledge unsuitable~\cite{chen2024toward}. MapReduce-based collaborative computing under energy-harvesting constraints has not yet been investigated.

{\color{black}Recent deep reinforcement learning (DRL)-based studies have considered long-term resource management in hierarchical federated learning~\cite{chen2024toward}, distributed unmanned aerial vehicle (UAV)-driven multi-access edge computing~\cite{li2025uav_mec}, and cellular-connected UAV path planning~\cite{li2022uav_qier}. Yet, these settings differ fundamentally from MapReduce-based collaborative computing, where computation, communication, phase timing, and energy dynamics are tightly coupled. 
We therefore develop a centralized DRL framework for long-term energy planning over heterogeneous energy-harvesting devices, while solving per-slot resource allocation via convex optimization.
Table~I compares the proposed framework with representative studies.}

\begin{table*}[t]
\centering
\caption{\color{black}Comparison with representative related studies.}
\label{tab:comparison_related_work}
\scriptsize
\renewcommand{\arraystretch}{1}
\setlength{\tabcolsep}{2.3pt}
\begin{tabular}{
>{\centering\arraybackslash}p{0.05\textwidth}
>{\RaggedRight\arraybackslash}p{0.17\textwidth}
>{\RaggedRight\arraybackslash}p{0.18\textwidth}
>{\RaggedRight\arraybackslash}p{0.18\textwidth}
>{\RaggedRight\arraybackslash}p{0.15\textwidth}
>{\RaggedRight\arraybackslash}p{0.19\textwidth}}
\hline
\textbf{Ref.} & \textbf{Supported model / assumptions} & \textbf{Main variables} & \textbf{Method} & \textbf{Complexity / overhead} & \textbf{Main distinction} \\
\hline
\cite{8051074} 
& Homogeneous MapReduce-type distributed computing 
& Computation load and communication load 
& Coded distributed computing with coded multicasting 
& Offline coding and data-placement design 
& No energy harvesting (EH) or online resource control \\
\hline
\cite{paris2021leveraging} 
& Collaborative fog computing with heterogeneous users 
& Computing load, CPU frequency, and transmit power 
& Convex optimization via KKT and Lagrange duality 
& Per-task centralized convex optimization 
& No long-term EH dynamics or DRL-based online planning \\
\hline
\cite{11005962,9795129,cheng2024asymptotically} 
& Heterogeneous coded distributed computing  
& Data placement, coding, and computation/communication load 
& MILP/approximation, or combinatorial design 
& Mainly offline MILP/coding/ design overhead 
& No joint phase-time, transmit-power, and EH battery control \\
\hline
\cite{chen2024toward} 
& Hierarchical federated learning
& Client scheduling and communication/computation resources 
& Two-phase DRL-based resource management 
& DRL training and policy inference 
& Different wireless control scenarios without MapReduce phase coupling \\
\hline
\cite{li2025uav_mec,li2022uav_qier} 
& UAV-driven MEC or cellular-connected UAV systems 
& UAV control, offloading, resource allocation, or path planning 
& Distributed many-agent DRL or QiER-enhanced DRL 
& DRL training and policy inference 
& Different wireless control scenarios without MapReduce phase coupling \\
\hline
\textbf{This work} 
& \textbf{MapReduce collaborative computing with heterogeneous EH devices} 
& \textbf{Load, phase time, transmit power, and energy budget} 
& \textbf{DDPG energy planning plus convex per-slot allocation} 
& \textbf{$O(KH^2)$ actor inference and $O(N^{3.5})$ convex solving} 
& \textbf{Joint MapReduce and EH-aware online control} \\
\hline
\end{tabular}
\end{table*}

The main contributions are summarized as follows.
\begin{enumerate}
  \item We establish a MapReduce-based framework for collaborative computing over
        heterogeneous wireless devices, in which the Map, Shuffle, and Reduce
        phases are explicitly modeled alongside computation energy, uplink
        transmission energy, and battery dynamics with energy harvesting, capturing the
        practical heterogeneity and energy sustainability requirements.

  \item We formulate a long-run average throughput maximization problem that
        jointly optimizes computing load, phase durations,
        transmit power, and device energy consumption under computation, communication, latency, and
        battery constraints. The formulation captures temporal coupling caused by battery dynamics and time-varying channels.

  \item We develop a two-phase algorithm in which Deep Deterministic Policy Gradient (DDPG) determines the per-device energy budgets based on battery and channel states, while convex optimization globally solves the per-slot resource allocation. This decomposition reduces the DRL action dimension, enabling faster convergence and higher throughput than DDPG alone.

\end{enumerate}
Simulations confirm that DDPG-CVX consistently achieves higher throughput
than all baselines across diverse battery and hardware configurations,
validating its effectiveness in heterogeneous, energy-constrained deployments.

The remainder of this paper is organized as follows. Section~II presents the system model and problem formulation.
Section~III details the DDPG-CVX algorithm. Section~IV evaluates performance,
and Section~V concludes the paper.

\section{System Model and Problem Formulation}\label{sec:model}
\subsection{MapReduce-Based Collaborative Computing}

\begin{figure}[t]
\centering
\includegraphics[scale=0.3]{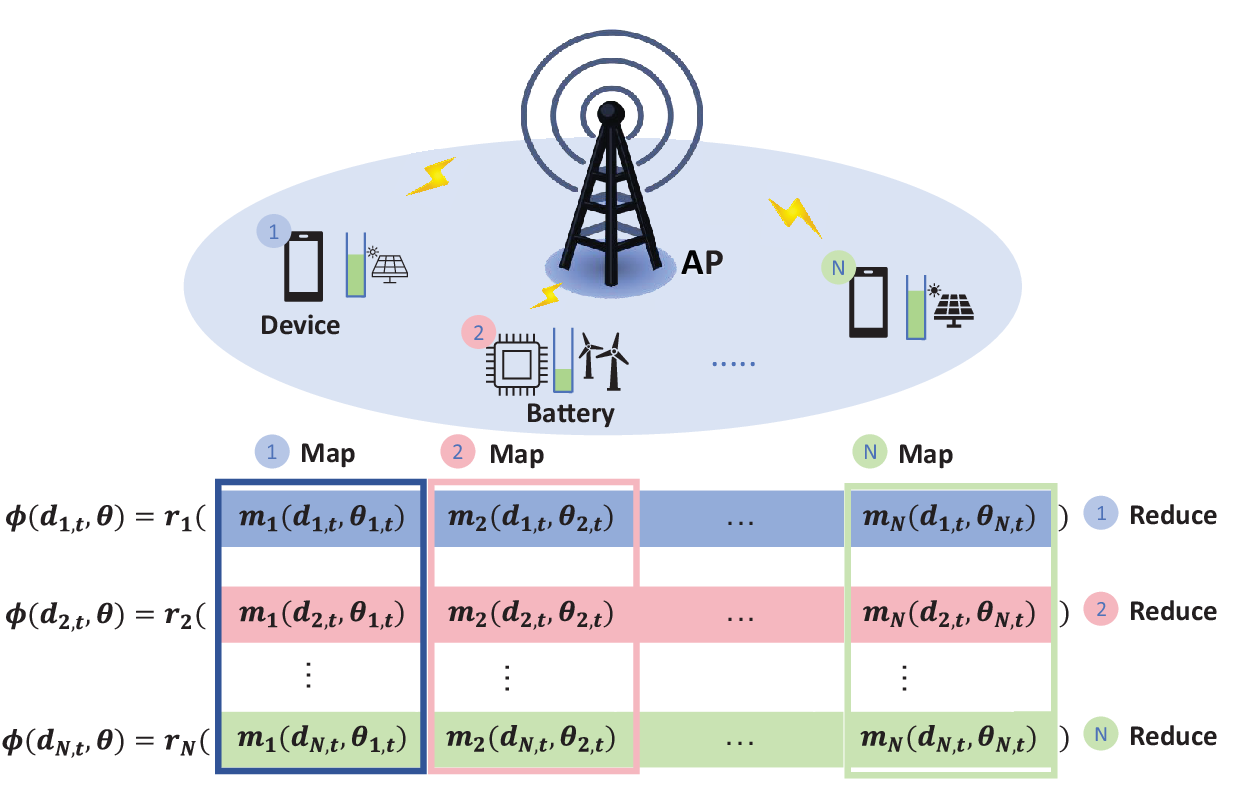}
\vspace{-2mm}
\caption{System model of MapReduce-based collaborative computing.}
\vspace{-4mm}
\label{fig:model}
\end{figure}

We consider $N$ heterogeneous wireless devices, indexed by $n\in\mathcal{N}:=\{1,\ldots,N\}$, connected to a common access point (AP). Time is divided into slots $t\in\mathcal{T}:=\{1,\ldots,T\}$ of duration $\tau$.
Each device $n$ aims to execute a specified computing task $\phi (d_{n,t}, {\bm{\theta}})$ within a time slot. Here, $d_{n,t}$ denotes the local input data for device $n$ at slot $t$, and ${\bm{\theta}}$ represents the $L$-bit data (e.g., a pretrained DL model) that is common to all $N$ devices. As illustrated in Fig.~\ref{fig:model}, $\phi (d_{n,t}, {\bm{\theta}})$ can represent the inference process performed using model ${\bm{\theta}}$ on input $d_{n,t}$ at slot $t$. Let $D^t$ denote the size of the input data $d_{n,t}$. Motivated by this application and the observation that DL models typically contain far more parameters than a single input sample, we assume $L \gg D^t$~\cite{8051074}. Due to the large size of ${\bm{\theta}}$, it may be infeasible for a single device to accomplish the computing task $\phi (d_{n,t}, {\bm{\theta}})$ within one time slot.

In the MapReduce framework, the $N$ devices collaborate to execute the computational tasks $\{\phi(d_{n,t},{\bm{\theta}})\}_{n=1}^N$. The $L$-bit data ${\bm{\theta}}$ can be partially processed at each time slot. Let $l_{n, t}$ denote the number of bits of ${\bm{\theta}}$ assigned to device $n$ at time slot $t$ (i.e., the size of partition ${\bm{\theta}}_{n, t}$), satisfying
\begin{equation}\label{l_lim}
\sum_{n=1}^{N} l_{n, t}\leq L.
\end{equation}

The local input data $\{d_{n,t}\}_{n=1}^N$ are assumed to have been shared among devices via the AP during initialization. The MapReduce framework consists of the following three phases.

\subsubsection{MAP}
In the Map phase, each device computes intermediate values based on its assigned DL model partition ${\bm{\theta}}_{n, t}$ of $l_{n, t}$ bits. The intermediate values computed by device $n$ (i.e., partial inference results using the model subset ${\bm{\theta}}_{n, t}$) are $\{m_n(d_{1,t},{\bm{\theta}}_{n, t}), m_n(d_{2,t},{\bm{\theta}}_{n, t}),\ldots, m_n(d_{n,t},{\bm{\theta}}_{n, t})\}$,
where $m_n$ denotes the Map function executed on device $n$.

\subsubsection{SHUFFLE}
In the Shuffle phase, the AP facilitates the exchange of intermediate computation results among devices. The size of the intermediate result $m_n(d_{x,t}, {\bm{\theta}}_{n, t})$ produced by device $n$ for device $x$ is proportional to the computing load (i.e., the size $l_{n,t}$ of ${\bm{\theta}}_{n,t}$), and is given by $\beta l_{n,t}$. Consequently, device $n$ must transmit $(N-1)\beta l_{n,t}$ bits to the other devices via the AP. For notational convenience, let $\alpha := (N-1) \beta$.


\subsubsection{REDUCE}
In the Reduce phase, each device $x$ aggregates a total of $\sum_{n=1}^N\beta l_{n,t}$ bits of intermediate computation results $\{m_n(d_{x,t}, {\bm{\theta}}_{n, t})\}_{n=1}^N$ received from all collaborating devices, and produces the final inference result of task $\phi (d_{x,t}, {\bm{\theta}})$ as $\{\rho_x\bigl(m_1(d_{x,t},{\bm{\theta}}_{1,t}), m_2(d_{x,t},{\bm{\theta}}_{2,t}),\ldots,m_N(d_{x,t},{\bm{\theta}}_{N,t})\bigr)\}$, where $\rho_x(\cdot)$ is the Reduce function executed at device $x$.

{\color{black}Let $t_{n,t}^\text{MAP}$, $t_{n,t}^\text{SHU}$, and $t_{n,t}^\text{RED}$ denote the latency of device $n$ in the Map, Shuffle, and Reduce phases during time slot $t$, respectively. Since Reduce begins only after all devices complete their Map and Shuffle operations, the system-level waiting time before Reduce is determined by the slowest device as
$T_t^{\text{MS}}=\max_{n\in\mathcal N}
\left(t_{n,t}^{\text{MAP}}+t_{n,t}^{\text{SHU}}\right)$.
To complete a MapReduce task within a single time slot, it must hold that
\begin{equation}\label{eq:tau}
T_t^{\text{MS}}+t_{n,t}^{\text{RED}}\leq \tau. 
\end{equation}
Since all devices prefer to complete the Reduce phase simultaneously at the end of slot $t$ to conserve energy, they share a common Reduce-phase latency $t_{n,t}^{\text{RED}}=t_t^{\text{RED}}$.}

\subsection{Computation,  Communication and Battery Models}

At time slot $t$, device $n$ processes the DL model partition ${\bm{\theta}}_{n, t}$ of $l_{n, t}$ bits for $N$ input samples during the Map phase. Let $c_n$ denote the effective number of CPU cycles required per bit of the model partition at device $n$, and let $k_{n}$ be the chip-architecture-dependent capacitance coefficient. The energy consumed for computation \cite{10498091} in the Map phase is
$E_{n, t}^\text{MAP}=\frac{k_{n}(Nc_{n} l_{n, t})^{3}}{\left(t_{n, t}^\text{MAP}\right)^{2}}$.
Similarly, the energy consumed by device $n$ to aggregate the intermediate results during the Reduce phase is
$E_{n, t}^\text{RED}=\frac{k_{n} (c_{n}\beta\sum_{n=1}^Nl_{n,t})^{3}}{\left(t_{n, t}^\text{RED}\right)^{2}}$.

Let $f_n^{\max}$ denote the maximum CPU frequency of device $n$. The CPU frequency constraints are
\begin{equation}\label{f1}
\frac{Nc_n l_{n,t} }{t_{n,t}^\text{MAP}}\leq  f_n^{\max}, ~~~
\frac{c_n \beta \sum_{n=1}^Nl_{n,t}}{t_{n,t}^\text{RED}} \leq  f_n^{\max}.
\end{equation}

During the Shuffle phase, devices exchange intermediate results through the AP using orthogonal frequency-division multiple access (OFDMA). Downlink transmission time and energy are neglected, as we focus on resource-constrained devices. Let $p_{n,t}\in[0,p_n^{\max}]$ denote the transmit power of device $n$ at slot $t$, where $p_n^{\max}$ is the maximum transmit power. According to Shannon's theorem, the achievable uplink rate of device $n$ at slot $t$ is $R_{n,t}(p_{n,t})=BW \log_2 \left(1+\frac{p_{n, t} h_{n, t}}{N_{0,t} BW}\right)$,
where $N_{0,t}$ is the noise power spectral density at the AP, ${h_{n,t}}$ is the channel gain from device $n$ to the AP, and $BW$ is the equally assigned uplink bandwidth for the devices.

{\color{black}Let $p_n^c$ denote the constant circuit power consumption at device $n$. The total energy consumed by device $n$ during the Shuffle phase is $E_{n,t}^\text{SHU}=p_n^c t_{n,t}^{\text{SHU}}+E_{n,t}^{\text{SHU,tx}}$, where $E_{n,t}^{\text{SHU,tx}}=p_{n,t}t_{n,t}^\text{SHU}$ is the transmission energy.} 
The data transmission constraint for the Shuffle phase is
\begin{equation}
\alpha l_{n,t} \leq t_{n,t}^\text{SHU} R_{n,t}\left(p_{n, t}\right),  \label{Ab}
\end{equation}
which ensures that the bits transmitted during the Shuffle phase are sufficient to deliver all required intermediate results.

{\color{black}Each device is powered by renewable energy and equipped with a rechargeable battery of capacity $E_{\max}$. Let $E_{n,t}^{\text{bat}}$ denote the battery level at the beginning of slot $t$. To reserve energy for sensing, signaling, and critical operations, we introduce a reserve level $E_{\text{res}}$ and impose
$E_{\text{res}} \leq E_{n,t}^{\text{bat}} \leq E_{\max},~\forall n,t.$ 
The battery state evolves as
\begin{equation}\label{battery}
E_{n,t+1}^{\text{bat}}=\min\!\left\{E_{n,t}^{\text{bat}}+E_{n,t}^{\text{get}}-E_{n,t}^{\text{tot}}, E_{\max}\right\}, 
\end{equation}
where $E_{n,t}^{\text{get}}$ is the stochastic renewable energy available to device $n$ at the beginning of slot $t$, harvested from solar, indoor light, or other ambient energy. $E_{n,t}^{\text{tot}} = E_{n,t}^\text{MAP} + E_{n,t}^\text{SHU} + E_{n,t}^\text{RED}$ is the total energy consumed by device $n$ at slot $t$.
The energy availability constraint holds to maintain the reserve level $E_{\text{res}}$: 
\begin{equation}\label{E_lim}
 E_{n,t}^{\text{tot}} \leq E_{n,t}^{\text{bat}}+E_{n,t}^{\text{get}}-E_{\text{res}}. 
\end{equation} }
\vspace{-6mm}
\subsection{Problem Formulation}
Let $\boldsymbol{\Phi}_t= \{l_{n,t}, t_{n,t}^\text{MAP},t_{n,t}^\text{SHU},t_{n,t}^\text{RED},p_{n,t}, E_{n,t}^{\text{tot}}, \forall n\} $. By optimizing $\boldsymbol{\Phi}_t$, the objective is to maximize the average system (computational) throughput. The problem is
\begin{equation}\label{P1}
\underset{\boldsymbol{\Phi}_t}{\operatorname{max}} \lim _{T \rightarrow \infty} \frac{1}{T} \sum_{t=1}^{T} \sum_{n=1}^{N} l_{n, t} \quad
\text{s.t.}~
\eqref{l_lim}\sim\eqref{E_lim},
 ~0 \le p_{n,t} \le p_n^{\max }.
\end{equation}
{\color{black}The $\lim_{T\to\infty}\frac{1}{T}\sum_{t=1}^T(\cdot)$ operator in \eqref{P1} defines the \emph{long-run time-average} throughput, a standard formulation for online resource allocation under stochastic, unpredictable channel and energy-arrival processes, as in \cite{chen2025energy}. In the numerical evaluation,
the long-run time-average objective is approximated by the finite-horizon empirical time-average $\frac{1}{T}\sum_{t=1}^T \sum_{n=1}^N l_{n,t}$ evaluated over a sufficiently large number of slots $T$.} 


\section{Proposed DDPG-CVX Framework}
Problem~\eqref{P1} is difficult to solve directly because future channel states and energy arrivals are unknown, while the battery dynamics in~\eqref{battery} couple decisions across time slots. Although problem~\eqref{P1} can be modeled as an Markov decision process (MDP), directly applying DDPG results in a high-dimensional action space that grows with the number of devices and hinders convergence.
We therefore propose a two-phase
framework.
{\color{black}In the first phase, DDPG determines the per-device energy budgets  $\bar{E}_{n,t}\geq E_{n,t}^{\text{tot}}$.} In the second phase, the remaining per-slot resource allocation is solved globally through convex optimization, and the resulting throughput is used as the reward for DDPG training.

\subsection{Proposed DDPG-Based Resource Control}

Problem~\eqref{P1} is formulated as an MDP with the 3-tuple $(\boldsymbol {S}, \boldsymbol {A}, \boldsymbol {r})$, comprising state $\boldsymbol {S}$, action $\boldsymbol {A}$, and reward $\boldsymbol {r}$.
The system state at time slot $t$ is $\boldsymbol {s}_t = \{ E_{n,t}^{\text{bat}},h_{n,t},\forall n\}\in\boldsymbol {S}$. 
The action at slot $t$ is $\boldsymbol {a}_t = \{\bar{E}_{n,t},\forall n\}\in \boldsymbol {A}$. 
{\color{black}Consistent with the objective in~\eqref{P1}, the immediate reward is defined as $r_t=\sum_{n=1}^N l_{n,t}-\zeta$, where $\zeta$ is a fixed positive penalty incurred when the DDPG-generated energy budget violates the energy availability constraint \eqref{E_lim}. This design avoids introducing an additional optimization bias.}

At slot $t$, the DDPG agent observes state $\boldsymbol{s}_t$, selects action $\boldsymbol{a}_t=\mu(\boldsymbol{s}_t)$, receives reward $r_t$, and transitions to $\boldsymbol{s}_{t+1}$. The action-value function is defined as
${Q^\mu }(\boldsymbol {s},\boldsymbol {a}) = {\mathbb{E}_\mu } [\sum\limits_{i  = 0}^\infty  {{\gamma ^i }r_{t + i}|{\boldsymbol {s}_t} = \boldsymbol {s},{\boldsymbol{a}_t} = \boldsymbol {a}} ]$, 
where $\gamma\in[0,1]$ is the discount factor. 
DDPG employs an actor network to approximate the policy $\mu$ and a critic network to estimate $Q^\mu(\boldsymbol{s},\boldsymbol{a})$. Each network has online and target versions, with the target networks improving training stability.



Let ${\mu}({\boldsymbol {s}_t}|{\boldsymbol{\omega} ^{\mu}})$ and ${Q}({\boldsymbol {s}_t},{\boldsymbol {a}_t}|{\boldsymbol{\omega} ^{{Q}}})$ denote the actor and critic online networks with parameters ${\boldsymbol{\omega} ^\mu }$ and ${\boldsymbol{\omega} ^Q}$, respectively. Let ${\mu'}({\boldsymbol {s}_t}|{\boldsymbol{\omega} ^{\mu'}})$ and ${Q'}({\boldsymbol {s}_t},{\boldsymbol {a}_t}|{\boldsymbol{\omega} ^{{Q'}}})$ denote the corresponding target networks with parameters ${\boldsymbol{\omega} ^{\mu'}}$ and ${\boldsymbol{\omega} ^{Q'}}$. The critic online network is updated by minimizing $\mathcal{L}({\boldsymbol{\omega} ^Q}) = \frac{1}{{{B}}}\sum\limits_t {[{{({\eta_t}-Q({\boldsymbol {s}_t},\mu ({\boldsymbol {s}_t|\boldsymbol{\omega}^ \mu})|{\boldsymbol{\omega} ^Q}) )}^2}]}$,
where ${\eta_t} = r_t + \gamma Q'({\boldsymbol {s}_{t+1}},\mu '({\boldsymbol {s}_{t + 1}|\boldsymbol{\omega} ^{\mu '}})|{\boldsymbol{\omega} ^{Q'}})$ and $B$ is the mini-\text{bat}ch size. The actor online network is optimized by ascending the gradient ${\nabla _{{\boldsymbol{\omega} ^\mu }}}J({\boldsymbol{\omega} ^\mu }) \approx \frac{1}{B}\sum\limits_t {{\nabla _{{\boldsymbol {a}_t}}}Q({\boldsymbol {s}_t},{\boldsymbol {a}_t}|{\boldsymbol{\omega} ^{\rm{Q}}})} {\nabla _{{\boldsymbol{\omega} ^\mu }}}\mu ({\boldsymbol {s}_t}|{\boldsymbol{\omega} ^\mu })$
to maximize the policy objective
$J({\boldsymbol{\omega} ^\mu }) = {\mathbb{E}_{{\boldsymbol{\omega} ^\mu }}}[Q({ \boldsymbol {s}_t},\mu ({\boldsymbol {s}_t}|{\boldsymbol{\omega} ^\mu })|{\boldsymbol{\omega} ^Q})]$, 
where ${\nabla _{{\boldsymbol{\omega} ^\mu }}}$ denotes the gradient with respect to $\boldsymbol{\omega} ^\mu$.
The target networks are updated via soft updates:
${\boldsymbol{\omega} ^{{\mu '}}} \leftarrow \lambda {\boldsymbol{\omega} ^\mu } + (1 - \lambda  ){\boldsymbol{\omega} ^{{\mu'}}}$ and ${\boldsymbol{\omega} ^{{Q'}}} \leftarrow \lambda  {\boldsymbol{\omega} ^Q} + (1 - \lambda ){\boldsymbol{\omega} ^{{Q'}}}$,
where $\lambda$ is the update rate.

\begin{figure}[t]
    \centering    \includegraphics[scale=0.26]{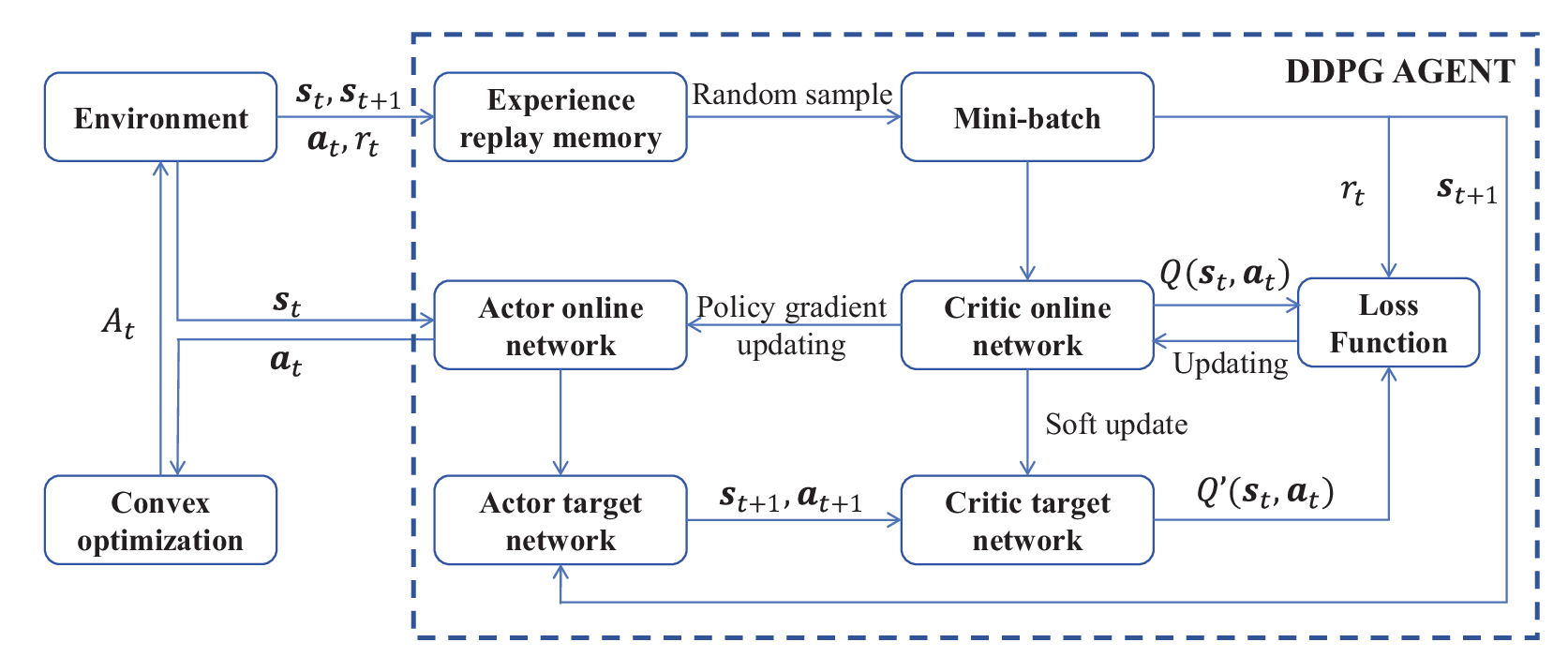}
    \vspace{-2mm}
    \caption{Architecture of the proposed DDPG-CVX framework.}
    \vspace{-4mm}
    \label{fig:DDPG_stucture}
\end{figure}

The DDPG-CVX workflow is illustrated in
Fig.~\ref{fig:DDPG_stucture}. At slot~$t$, the actor generates the
energy budgets $\boldsymbol{a}_t$ from the observed state
$\boldsymbol{s}_t$. Given $\boldsymbol{a}_t$, the remaining optimization variables $\boldsymbol{\Phi}'_t:=\{t_{n,t}^\text{MAP},t_{n,t}^\text{SHU},t_{t}^\text{RED},p_{n,t},l_{n,t}, \forall n, t\}$ are determined via convex optimization to evaluate the reward, as detailed in Section~\ref{sec:cvx}. The slot concludes with the agent receiving reward $r_t$ and transitioning to state $\boldsymbol {s}_{t+1}$. {\color{black}During training, zero-mean Gaussian noise is added to the actor output for exploration, and the resulting energy budgets are clipped to the feasible range.} 
Transitions
$\{\boldsymbol{s}_t,\boldsymbol{a}_t,r_t,\boldsymbol{s}_{t+1}\}$
are stored in an experience replay buffer. Mini-batches of $B$
transitions are sampled to update the actor and critic online networks, followed by the target-network updates. Experience replay reduces temporal correlations
among training samples and improves learning stability.

{\color{black}The MDP assumes that $E_{n,t}^{\text{bat}}$ and $h_{n,t}$ are available at the decision time. In practice, only the estimated state $\hat{\boldsymbol{s}}_t=\{\hat{E}_{n,t}^{\text{bat}},\hat{h}_{n,t},\forall n\}$ may be observed because of estimation errors or delayed feedback. Exploration and experience replay of DDPG can improve tolerance to mild errors, while stronger uncertainty can be addressed through noisy-state training, history-augmented policies, or conservative energy margins.}


\subsection{Resource Allocation via Convex Optimization}
\label{sec:cvx}

{\color{black}At each time slot $t$, given the action $\boldsymbol{a}_t = 
\{\bar{E}_{n,t}, \forall n\}$ generated by DDPG, 
problem~\eqref{P1} reduces to the following per-slot subproblem 
without battery constraints:
\begin{align}\label{Ua}
\underset{\boldsymbol{\Phi}'_t}{\operatorname{max}} 
&\sum_{n=1}^{N} l_{n, t} \\\notag  
\text{s.t.} \quad  
&\eqref{l_lim}\sim\eqref{Ab},\,0 \le p_{n,t} \le p_n^{\max },\\\notag
&E_{n,t}^{\text{tot}}=E_{n,t}^\text{MAP}+E_{n,t}^\text{SHU}+E_{n,t}^\text{RED}
\leq \bar{E}_{n,t}, \quad \forall n.
\end{align}}%
The linear constraints, including the time slot constraint 
$t_{n,t}^\text{MAP}+t_{n,t}^\text{SHU}+t_{t}^\text{RED}\leq\tau$ 
and the CPU frequency constraints 
$Nc_n l_{n,t} \leq f_n^{\text{max}} t_{n,t}^{\text{MAP}}$ and 
$c_n\beta \sum_nl_{n,t} \leq f_n^{\text{max}} t_{n,t}^{\text{RED}}$, 
trivially satisfy convexity.
{\color{black}For the communication constraint 
$\alpha l_{n,t}\leq t_{n,t}^{\text{SHU}} R_{n,t}(p_{n,t})$, 
we introduce the substitution 
$E_{n,t}^{\text{SHU,tx}}=p_{n,t}t_{n,t}^{\text{SHU}}$ to eliminate $p_{n,t}$, yielding
\begin{equation}\label{16}
\alpha l_{n,t}
-
t_{n,t}^{\text{SHU}} BW
\log_2\left(
1+
\frac{E_{n,t}^{\text{SHU,tx}}h_{n,t}}
{N_0 BW t_{n,t}^{\text{SHU}}}
\right)
\leq 0,\quad \forall n,t. 
\end{equation}
The left-hand side of \eqref{16} is convex in $(l_{n,t}, t_{n,t}^{\text{SHU}}, E_{n,t}^{\text{SHU,tx}})$. 
The term $\alpha l_{n,t}$ is linear, and 
the second term is concave in 
$(t_{n,t}^{\text{SHU}},\,E_{n,t}^{\text{SHU,tx}})$ as the 
perspective of the concave logarithmic function~\cite{convex}, 
so their difference is convex.} 
Additionally,  
$E_{n,t}^{\text{MAP}} = k_n (Nc_n l_{n,t})^3 / 
(t_{n,t}^{\text{MAP}})^2$ is jointly convex in 
$(l_{n,t},\, t_{n,t}^{\text{MAP}})$ for $l_{n,t}, 
t_{n,t}^{\text{MAP}} > 0$, as it takes the form 
$x^3/y^2$ whose positive semi-definiteness of the 
Hessian can be verified directly. Similarly, the Reduce phase energy
$E_{n,t}^{\text{RED}} = k_n (c_n \beta \sum_nl_{n,t})^3 / 
(t_{n,t}^{\text{RED}})^2$ is convex in $(l_{n,t},t_{n,t}^{\text{RED}})$.
{\color{black}Since the objective and all constraints are convex in $\boldsymbol{\Phi}'_t$, subproblem~\eqref{Ua} is a convex program and can be solved globally using standard methods such as the interior-point method~\cite{convex}. Once DDPG fixes the energy budgets $\{\bar{E}_{n,t}\}$, CVX optimizes the remaining computing load, phase durations, and transmit energy. This yields an accurate reward signal and avoids the infeasibility caused by directly exploring all resource variables with DRL.

\textbf{Complexity Analysis.} The per-slot complexity consists of actor inference and convex optimization. For a $K$-layer fully connected actor with hidden width $H$, the inference complexity is approximately $\mathcal{O}(KH^2)$. The convex subproblem~\eqref{Ua} contains $(4N+1)$ variables and has a worst-case interior-point complexity of $\mathcal{O}(N^{3.5})$, which is tractable for moderate-scale deployments (e.g., up to a few tens of devices per AP, consistent with typical MapReduce
collaborative-computing group sizes). For larger systems, computation can be reduced using warm starts, first-order solvers, or neural surrogates, which are left for future work.}


\section{Performance Evaluation}
{\color{black}We set $N=5$, $BW=30$~MHz,  $N_{0,t}=1$~nW/Hz, and $\tau=100$~ms. Channel gains follow a complex Gaussian
distribution $\mathcal{CN}(0,10^{-3})$. Device hardware parameters are
drawn heterogeneously: $E_{\max}\in [100,10,000]~\mu\mathrm{J}$, $k_n \in [10^{-28},
10^{-27}]$, $p_n^{\max} \in [10, 25]$~mW, $p_n^c \in [10, 25]$~mW, $f_n^{\max} \in [1,
3]$~GHz, and $c_n \in [500, 1500]$. The harvested energy is generated independently for each device and time
slot following a bounded uniform distribution $E_{n,t}^{\text{get}} \sim \mathcal{U}[0,200]~\mu\mathrm{J}$. 
Both the actor and critic networks contain two hidden fully connected layers with 32 and 64 neurons, respectively, and ReLU activations.
The discount factor is $\gamma=0.95$, the actor/critic learning rate is $2\times10^{-3}$, the replay-buffer size is 4000, $B=32$, and $\lambda=0.01$.}

















{\color{black}We compare the proposed DDPG-CVX algorithm against the following baselines: \textbf{1) Greedy Allocation (GA):} Each device depletes its entire battery at every time slot, i.e., $\bar{E}_{n,t}=E_{n,t}^{\text{bat}}$. \textbf{2) Random Budget (RB):} The energy budgets $\bar{E}_{n,t}$ are randomly generated within the feasible range.
\textbf{3) Average Allocation (AA):} The computational load is divided equally among all devices, i.e., $l_{n,t}=l_{m,t}, \forall n, m\in\mathcal{N}$.
\textbf{4) DDPG-only:} The full optimization problem is solved directly by a standalone DDPG agent.
\textbf{5) MAX Frequency (MAXF):} Every device operates at its peak CPU frequency $f_n^{\max}$ regardless of the energy state. For GA, RB, AA and MAXF, the remaining variables are obtained using CVX. For DDPG-only, the normalized actor outputs are mapped to physical resource variables and projected onto the feasible set. Its best-performing configuration uses actor and critic learning rates of $10^{-3}$, a replay-buffer capacity of $10^5$, a mini-batch size of 64, $\gamma=0.95$, and zero-mean Gaussian exploration noise.}

\begin{figure}[t]
  \centering
  \includegraphics[scale=0.25]{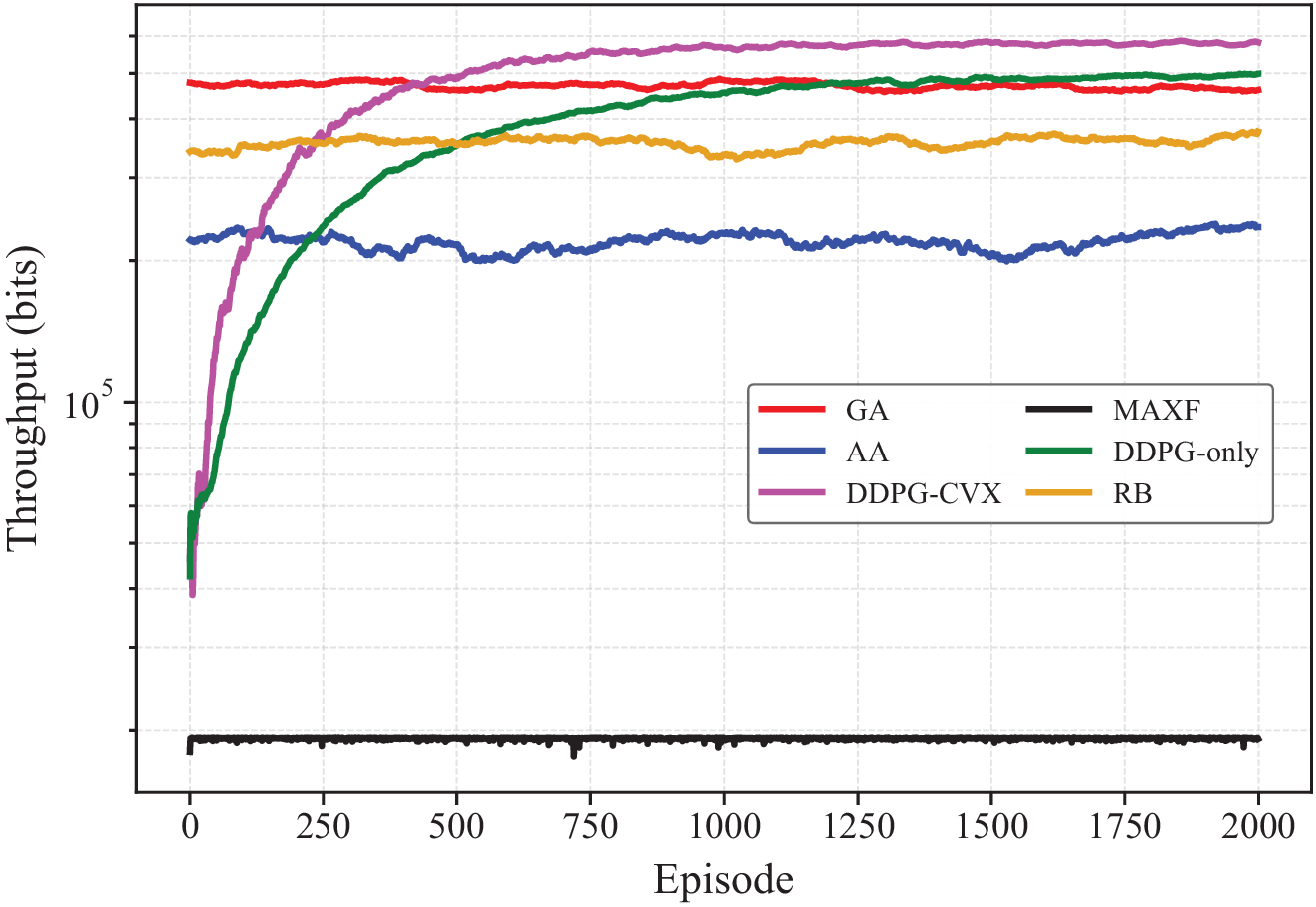}
  \vspace{-3mm}
  \caption{Throughput performance over episodes under different algorithms.}
  \vspace{-2mm}
  \label{fig:throughput_convergence}
\end{figure}

{\color{black}Fig.~\ref{fig:throughput_convergence} shows the throughput performance over episodes. DDPG-CVX converges to the highest
throughput within 1,000 episodes, while DDPG-only converges more
slowly and to a lower value due to the high-dimensional action space.
GA, RB and AA settle at substantially lower levels, as their fixed or random policies cannot adapt to varying channel and battery conditions.
MAXF performs worst, as continuous peak-frequency operation rapidly depletes the batteries and limits communication.
Specifically, DDPG-CVX achieves 
$1.25\times$, $1.51\times$, $1.75\times$, $2.87\times$, and $32.36\times$
the throughput of DDPG-only, GA, RB, AA, and MAXF, respectively.}  

\begin{figure}[t]
  \centering
  \includegraphics[scale=0.25]{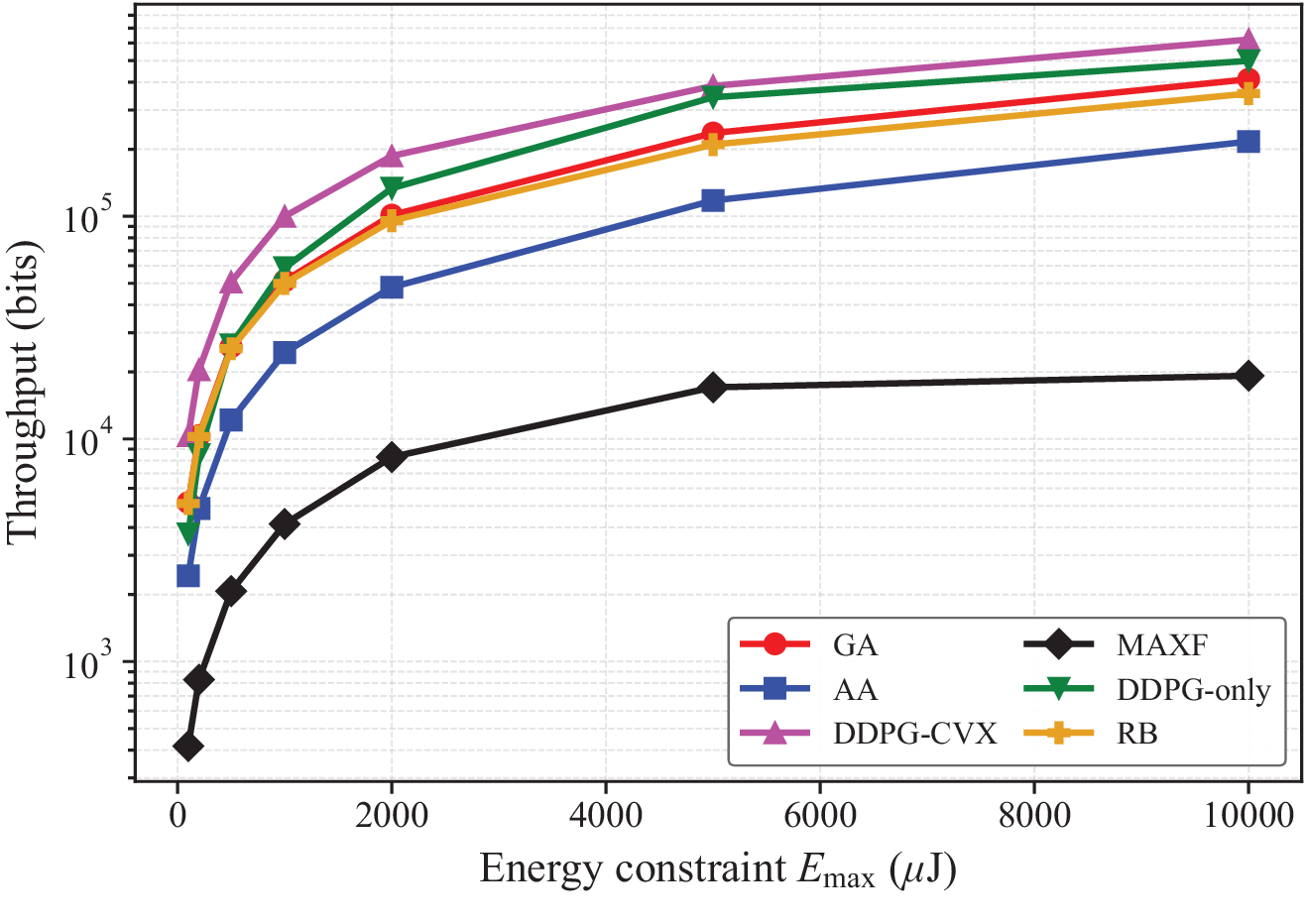}
  \vspace{-3mm}
  \caption{System throughput versus different battery capacities.}
  \vspace{-5mm}\label{fig:throughput_energy}
\end{figure}

Fig.~\ref{fig:throughput_energy} shows the system throughput versus battery capacity $E_{\max}$, ranging from $100$ to $10{,}000~\mu$J. DDPG-CVX achieves the highest throughput across the entire range, as larger batteries enable sustained computation and the proposed method jointly optimizes computing load, transmit power, and phase durations. Its advantage over DDPG-only demonstrates the benefit of solving the per-slot resource allocation through convex optimization rather than exploring all variables directly in a high-dimensional action space.


\begin{figure}[t]
  \centering
  \includegraphics[scale=0.25]{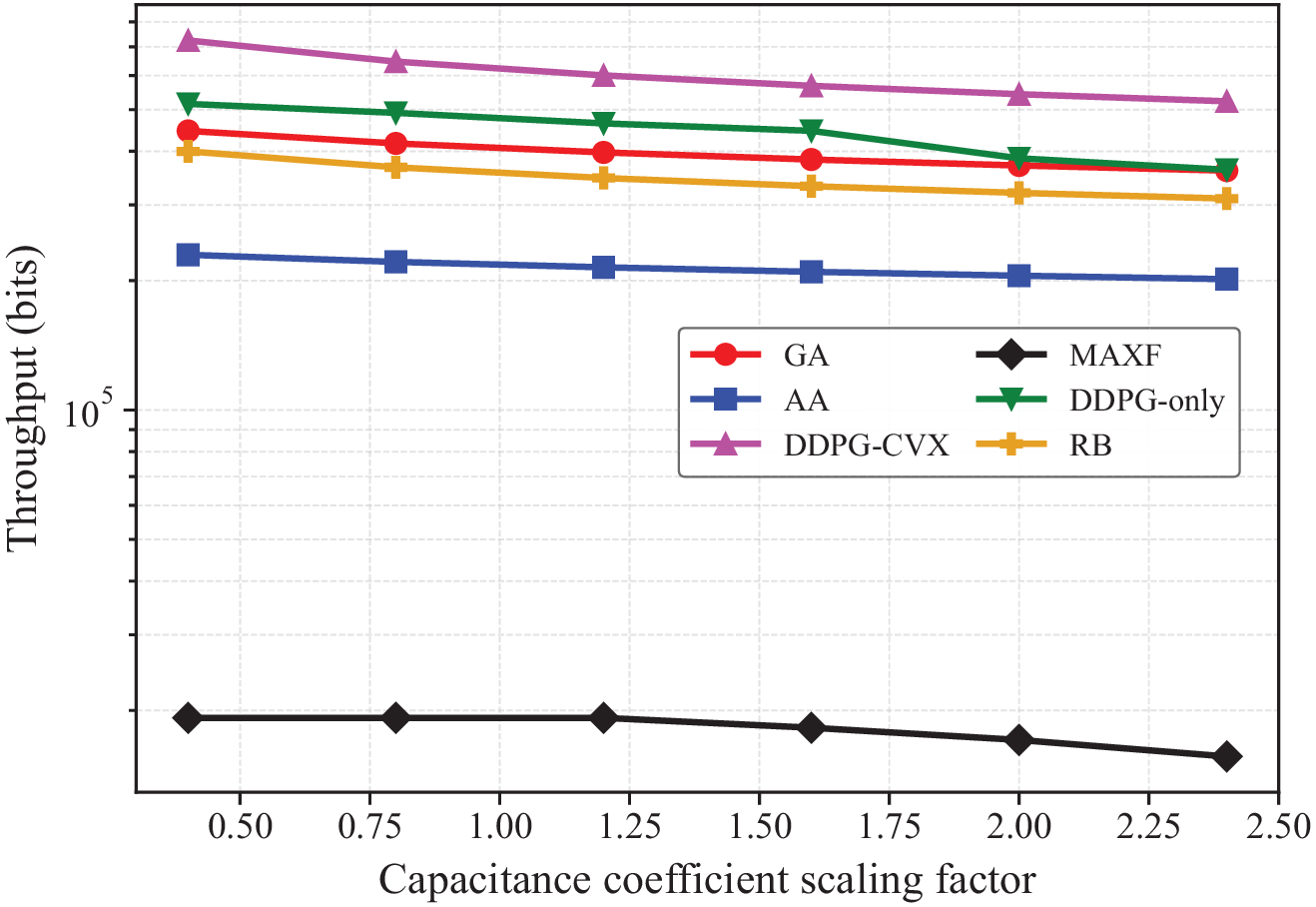}
  \vspace{-3mm}
  \caption{System throughput versus different capacitance coefficients.}
  \vspace{-3mm}
  \label{fig:throughput_kn}
\end{figure}

Fig.~\ref{fig:throughput_kn} examines robustness to hardware heterogeneity
by scaling the capacitance coefficient $k_n$ from $0.4$ to $2.4$. As $k_n$
grows, all algorithms suffer throughput loss since each CPU cycle becomes more
energy-intensive, constraining the data volume processed per slot. Despite this uniform
trend, DDPG-CVX maintains the highest throughput across the entire range by
jointly redistributing CPU frequencies and computing load to compensate for
the increased per-cycle energy cost. 

\begin{table}[t]
\centering
\caption{\color{black}Throughput under different hyperparameter settings.}
\label{tab:hyperparameter_sensitivity}
\renewcommand{\arraystretch}{1}
\scriptsize
\setlength{\tabcolsep}{15pt}
\begin{tabular}{lccc}
\toprule
\multirow{2}{*}{Learning rate}
& \multicolumn{3}{c}{Discount factor $\gamma$} \\
\cmidrule(lr){2-4}
& $0.90$ & $0.95$ & $0.99$ \\
\midrule
$10^{-3}$& 188192& 186158& 182387\\
$2\times10^{-3}$& 187069& \textbf{189218}& 186401\\
$5\times10^{-3}$& 184445& 184865& 185268\\
\bottomrule
\end{tabular}
 \vspace{-5mm}
\end{table}

{\color{black}We further test the discount factor $\gamma\in\{0.90,0.95,0.99\}$ and the actor/critic learning rate
$\eta_a=\eta_c\in\{10^{-3},2\times10^{-3},5\times10^{-3}\}$. 
As shown in Table~\ref{tab:hyperparameter_sensitivity}, 
$\gamma=0.95$ and $\eta_a=\eta_c=2\times10^{-3}$ achieve the highest throughput among the tested settings. 
The results show that the achieved throughput is sensitive to the discount factor and learning-rate settings, highlighting the need for careful hyperparameter tuning.} 

\section{Conclusion}
\label{sec:con}
This paper proposed DDPG-CVX for MapReduce-based collaborative computing over heterogeneous energy-harvesting devices. DDPG performs long-term energy planning, while convex optimization globally solves the per-slot resource allocation, reducing the DRL action dimension. Simulation results demonstrate consistent throughput gains over the considered baselines under diverse hardware configurations.


\bibliographystyle{IEEEtran}
\bibliography{references}
\end{document}